\documentclass[prb,twocolumn,aps,superscriptaddress,showpacs,amsmath,amssymb]{revtex4}
\usepackage{dcolumn}

\usepackage{amsmath,amssymb,graphicx,bm,xcolor}

\begin{document}

\title{Legendre Transformation of the Luttinger--Ward Functional\\ from the Bare Interaction Vertex to the Renormalized One}
\author{Takafumi K{\sc ita}}
\affiliation{Department of Physics, Hokkaido University, Sapporo 060-0810, Japan}

\begin{abstract}
On the basis of the Luttinger--Ward functional for interacting many-body systems
given in terms of full Green's function $G$ and the bare interaction vertex $\Gamma^{(0)}$, 
we develop a novel Legendre transformation 
to express the grand thermodynamic potential $\Omega$ as a functional of  $G$
and the renormalized interaction vertex $\Gamma$ 
so that (i) $G$ and $\Gamma$ obey the stationarity conditions $\delta \Omega/\delta G=0$ and $\delta\Omega /\delta\Gamma=0$ and (ii)  $\Gamma$ reduces to  $\Gamma^{(0)}$ in the weak-coupling limit.
The formalism enables us to perform microscopic studies of thermodynamic, 
single-particle, and two-particle properties in a unified self-consistent conserving framework.
\end{abstract}

\maketitle

\section{Introduction}

The purpose of the present study is to express the grand thermodynamic potential $\Omega$ 
as a functional of full Green's function $G$ and the renormalized interaction vertex $\Gamma$
to incorporate single-particle excitations and collective fluctuations microscopically into thermodynamics.
So far, the vertex renormalization has been performed mostly by procedures with no explicit connection to $\Omega$,
such as the $T$-matrix or shielded-potential approximation \cite{BK61,Baym62,BS89},
but we then encounter a difficulty in taking the resulting vertex into $\Omega$.
Expressing $\Omega$ in terms of $\Gamma$ is prerequisite to quantitative calculations
of thermodynamics with two-particle or collective excitations, especially around emerging ordered phases
where those fluctuations become dominant.  

The topic was pioneered by De Dominicis and Martin \cite{DDM64-1,DDM64-2} 
who introduced the Legendre transformation of $\Omega$ as a key factor 
in performing the renormalizations at various levels.
Using this transformation, they presented a diagrammatic vertex renormalization procedure \cite{DDM64-2} that relies on the topological structure of skeleton-type Feynman diagrams for $\Omega$. 
However, no explicit proof has been given on the validity of assigning the same sign and weight as those of the bare perturbation expansion
to each of the renormalized skeleton-type diagrams. 
Moreover, the resulting $\Gamma$ has an unfavorable feature that it reduces to $-\Gamma^{(0)}$ 
instead of  the bare interaction vertex $\Gamma^{(0)}$ itself in the weak-coupling limit; see Eq.\ (52) of Ref.\ \onlinecite{DDM64-2} on this point with $\Gamma^{(0)}\rightarrow v_2$ and $\Gamma\rightarrow C_2$.
Among other approaches to incorporating collective excitations into thermodynamics is that based on the Hubbard--Stratonovich transformation \cite{Stratotovich57,Hubbard59}. However, it generally focuses on a single kind of fluctuation such as the density or spin 
upon performing the transformation so that the other fluctuations may be neglected completely in any approximate treatment afterwards.

With these as a background, in this study we develop an alternative vertex renormalization procedure 
based on the Luttinger--Ward functional \cite{LW60},
where the renormalization at the one-particle level, called ``mass renormalization'', has been completed rigorously through expressing $\Omega$
as a functional of $G$.
Since it is written in terms of $\Gamma^{(0)}$, however,
the functional still has room for a further renormalization.
A key quantity in it is the functional $\Phi[G,\Gamma^{(0)}]$ composed of skeleton diagrams for $\Omega$, 
which forms a basis for systematic approximations called {\it $\varPhi$-derivable}\cite{Baym62} or  {\it conserving}\cite{BS89} {\it approximation} that enable us to incorporate correlation effects progressively
beyond the Hartree--Fock theory to describe equilibrium and nonequilibrium many-body phenomena\cite{LW60,BK61,Baym62,BS89,Kita10}.
Using the correlation part of $\Phi[G,\Gamma^{(0)}]$, we will perform the remaining renormalization
to express $\Omega$ as a functional of $(G,\Gamma)$ in such a way that the functional obeys the stationarity conditions $\delta \Omega/\delta G=0$ and $\delta\Omega/\delta\Gamma=0$.
The latter condition yields a self-consistent equation for $\Gamma$
whose solution appropriately reduces to $\Gamma^{(0)}$ in the weak-coupling limit.
Once $G$ and $\Gamma$ are known by solving the coupled equations, 
we can also obtain the two-particle Green's function $G^{\rm II}$, as will be shown below.
Thus, the present formalism provides a unified framework to study $(\Omega,G,G^{\rm II})$ consistently and simultaneously.
For example, it enables us to calculate free energy, magnetization, and susceptibility
through magnetic transitions, 
incorporating single-particle and collective excitations on an equal footing.

This paper is organized as follows. In Sect.\ 2, we present the formulation. In Sect.\ 3, we provide several examples of approximate $\Phi$ functionals for fermions.
In Sect.\ 4, we provide concluding remarks. We use the units of $\hbar=k_{\rm B}=1$.
 
\section{Formulation}

\subsection{Luttinger--Ward functional}

We consider a system of identical bosons or fermions interacting via a two-body potential $U({\bm r}_1-{\bm r}_2)$
in the coordinate-space representation.
We focus on normal states for clarity, 
for which the Luttinger--Ward functional is given by \cite{LW60,Kita10} 
\begin{align}
\Omega=\frac{\sigma}{\beta}{\rm Tr}\bigl[\ln(-\underline{G}_0^{-1}+\underline{\Sigma})+\underline{\Sigma}\,\underline{G}\bigr]+\Phi.
\label{Omega_LW}
\end{align}
Here, $\sigma$ takes $+1$ for bosons and $-1$ for fermions, $\beta$ is the inverse temperature, 
and $\underline{G}_0$ and $\underline{\Sigma}$ respectively are noninteracting matrix Green's function and self-energy whose rows and columns are specified by the index $1\equiv (\xi_1,\tau_1)$, where $\xi_1$ denotes the space-spin coordinates\cite{Kita15} and 
$\tau_1$ is the imaginary time that lies in $[0,\beta]$.
This functional satisfies the stationarity condition $\delta \Omega/\delta \underline{G}=0$,
which yields Dyson's equation\cite{Kita10}
\begin{align}
\underline{G}^{-1}=\underline{G}_0^{-1}-\underline{\Sigma},
\label{Dyson}
\end{align}
where the self-energy is defined by
\begin{align}
\Sigma(1',1)=-\sigma\beta \frac{\delta\Phi}{\delta G(1,1')} .
\label{Sigma}
\end{align}
\begin{figure}[t]
\begin{center}
\includegraphics[width=0.45\textwidth]{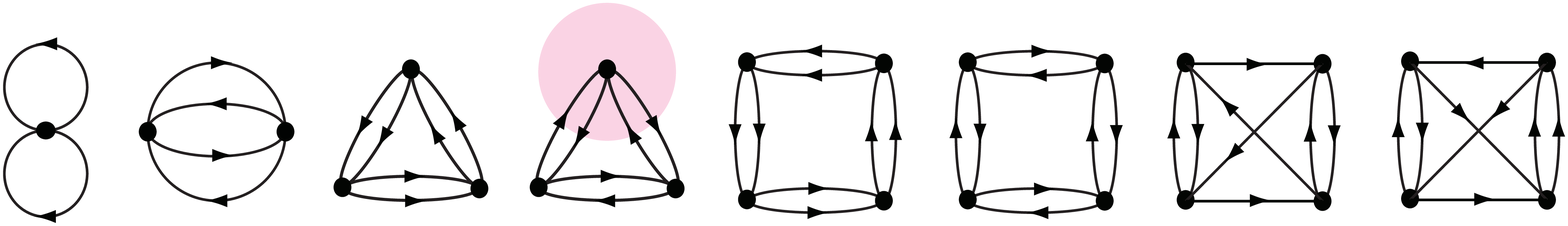}
\caption{(Color online) Diagrammatic expression of  $\Phi$ up to the fourth order in  $\Gamma^{(0)}$. A filled circle and a line with an arrow denote $\Gamma^{(0)}$ and $G$, respectively. The area enclosed by the circle in the fourth diagram represents an extended vertex $\bar{\Gamma}^{(0)}$ defined by Eq.\ (\ref{barGamma^(0)}).
\label{Fig1}}
\end{center}
\end{figure}
A key quantity in Eq.\ (\ref{Omega_LW}) is the functional $\Phi$. To write it concisely, we introduce the bare symmetrized vertex,
\begin{align}
\Gamma^{(0)}(1'2',12)
\equiv &\,U({\bm r}_1-{\bm r}_2)\delta(\tau_1-\tau_2)\bigl[\delta(1',1)\delta(2',2)
\notag \\
&\,+\sigma
\delta(1',2)\delta(2',1)\bigr] ,
\label{Gamma^(0)}
\end{align}
having the symmetry $\Gamma^{(0)}(1'2',12)\!=\!\sigma \Gamma^{(0)}(2'1',12)\!=\!\sigma \Gamma^{(0)}(1'2',21)
\!=\!\Gamma^{(0)}(12,1'2')$.
 The interaction Hamiltonian can be expressed in terms of $\Gamma^{(0)}$ and 
 the field operators $(\hat{\psi},\hat{\psi}^\dagger)$ in the Heisenberg representation as
\begin{align}
\hat{H}_{\rm int}=&\,\frac{1}{4\beta}\hat{\psi}^\dagger(1')\hat{\psi}^\dagger(2')\Gamma^{(0)}(1'2',12)\hat{\psi}(2)\hat{\psi}(1) ,
\label{H_int-1}
\end{align}
where integrations over repeated arguments are implied.
As given in Fig.\ 1 graphically, the functional $\Phi$ consists of skeleton diagrams in the bare perturbation expansion of $\Omega$ 
in $\hat{H}_{\rm int}$ with the replacement 
$\underline{G}_0\rightarrow \underline{G}$.
The first-order diagram is the Hartree--Fock or mean-field contribution given analytically by
\begin{align}
\Phi_{\rm MF}\equiv &\, \frac{1}{2\beta} {\Gamma}^{(0)}(1'2',12)G(1,1_+')G(2,2_+'),
\end{align}
where the subscript $_+$ 
 in the equal-time averages of the first order
places the creation operator to the left.\cite{LW60,Kita10}

The remaining diagrams in Fig.\ \ref{Fig1} constitute the correlation contribution  $\Phi_{\rm c}\equiv \Phi-\Phi_{\rm MF}$. 
To express $\Phi_{\rm c}$ concisely, we follow the procedure of De Dominicis and Martin \cite{DDM64-1} to decompose
matrix Green's function formally as $\underline{G}=\underline{G}^{\frac{1}{2}}\underline{G}^{\frac{1}{2}}$ and introduce
the extended vertex,
\begin{align}
\bar{\Gamma}^{(0)}(1'2',12)\equiv &\, G^{\frac{1}{2}}(1',3')G^{\frac{1}{2}}(2',4')\Gamma^{(0)}(3'4',34)
\notag \\
&\,\times
G^{\frac{1}{2}}(3,1)G^{\frac{1}{2}}(4,2) ,
\label{barGamma^(0)}
\end{align}
which represents the structure enclosed by the circle in Fig.\ \ref{Fig1} with
the first two (latter two) arguments corresponding to outgoing (incoming) lines.
Indeed, we can write $\Phi_{\rm c}$ as a functional of $\bar{\Gamma}^{(0)}$ alone as
\begin{align}
\Phi_{\rm c}\bigl[\bar{\Gamma}^{(0)}\bigr]=&\,-\frac{1}{8\beta}\bar{\Gamma}^{(0)}(1'2',12)\bar{\Gamma}^{(0)}(12,1'2')
\notag \\
&\,\hspace{-10mm}+\frac{\sigma^2}{2^3\cdot 3\beta}\bar{\Gamma}^{(0)}(11',22')\bar{\Gamma}^{(0)}(22',33')\bar{\Gamma}^{(0)}(33',11')
\notag \\
&\,\hspace{-10mm}+\frac{\sigma^3}{2\cdot 3\beta}\bar{\Gamma}^{(0)}(1'2,12')\bar{\Gamma}^{(0)}(2'3,23')\bar{\Gamma}^{(0)}(3'1,31')
\notag \\
&\,\hspace{-10mm}-\cdots .
\label{Phi_c}
\end{align}
The prefactor of the second term on the right-hand side originates from the product of (i) $-\beta^{-1}$ in the definition of $\Omega$, 
(ii) $(-\frac{1}{4})^3\frac{1}{3!}$ from the expansion of $e^{-\beta\hat{H}_{\rm int}}$, (iii) $2! 2^3$ from the 
number of ways of connecting the creation and annihilation operators for the third diagram in Fig.\ 1, and (iv) $\sigma^2$ from the two closed particle lines.
That of the third term has been obtained by replacing the latter two factors above by (iii) $2! 2^5$ and (iv) $\sigma^3$, respectively,
as appropriate for the fourth diagram in Fig.\ 1.
Specifically on (iii), every factor $2$ originates from choosing one of the two identical creation or annihilation operators in the interaction Hamiltonian  to connect.

\subsection{Renormalized vertex and Legendre transformation}

Now, we perform the Legendre transformation for changing the vertex variable in $\Omega$ from $\Gamma^{(0)}$ to $\Gamma$.
First of all, we define the renormalized vertex in terms of $\Phi_{\rm c}$ in Eq.\ (\ref{Phi_c}) by
\begin{align}
\bar\Gamma(1'2',12)\equiv &\, -4\beta\frac{\delta \Phi_{\rm c}}{\delta \bar\Gamma^{(0)}(12,1'2')}.
\label{bGamma-def}
\end{align}
Practically, we should replace $\bar\Gamma^{(0)}(12,1'2')$ in the denominator by $\frac{1}{2}[\bar\Gamma^{(0)}(12,1'2')+\sigma\bar\Gamma^{(0)}(12,2'1')]$ to incorporate the symmetry of $\bar\Gamma^{(0)}$ manifestly in the differentiation.
Substitution of Eq.\ (\ref{Phi_c}) into the right-hand side yields
\begin{align}
\bar\Gamma(1'2',12)
=&\, \bar{\Gamma}^{(0)}(1'2',12)
-\frac{1}{2}\bar{\Gamma}^{(0)}(1'2',33')\bar{\Gamma}^{(0)}(33',12)
\notag \\
&\,\hspace{-10mm}- \sigma\bigl[\bar{\Gamma}^{(0)}(2'3,23')\bar{\Gamma}^{(0)}(3'1',31)
\notag \\
&\,\hspace{-2mm}+\sigma\bar{\Gamma}^{(0)}(1'3,23')\bar{\Gamma}^{(0)}(3'2',31)\bigr]+\cdots ,
\label{bGamma-bGamma^(0)}
\end{align}
which certainly reduces to $\bar\Gamma^{(0)}$ in the weak-coupling limit.
Equation (\ref{bGamma-bGamma^(0)}) enables us to express $\bar{\Gamma}^{(0)}$ as a functional of $\bar\Gamma$.
We then perform the Legendre transformation,
\begin{align}
\Phi_{\rm L}\bigl[\bar\Gamma\bigr]\equiv \frac{1}{4\beta}\bar\Gamma^{(0)}(1'2',12)\bar\Gamma(12,1'2')+\Phi_{\rm c}\bigl[\bar\Gamma^{(0)}\bigr],
\label{Phi_L}
\end{align}
which satisfies
\begin{align}
4\beta \frac{\delta \Phi_{\rm L}\bigl[\bar\Gamma\bigr]}{\delta \bar\Gamma(12,1'2')}=\bar\Gamma^{(0)}(1'2',12),
\label{dbarPhi/dGamma}
\end{align}
since the implicit dependence on $\bar\Gamma$ through $\bar\Gamma^{(0)}$
cancels out owing to Eq.\ (\ref{bGamma-def}).
Equation (\ref{dbarPhi/dGamma}) is expressible as a stationarity condition in terms of $\bar\Gamma$
by introducing another functional of $(\bar\Gamma^{(0)},\bar\Gamma)$,
\begin{align}
\bar\Phi_{\rm c}\bigl[\bar\Gamma^{(0)},\bar\Gamma\bigr]\equiv -\frac{1}{4\beta}\bar\Gamma^{(0)}(1'2',12)\bar\Gamma(12,1'2')
+\Phi_{\rm L}\bigl[\bar\Gamma\bigr].
\label{barPhi_c}
\end{align}
Indeed, one can show by using Eq.\ (\ref{dbarPhi/dGamma}) that
\begin{align}
\frac{\delta\bar\Phi_{\rm c}\bigl[\bar\Gamma^{(0)},\bar\Gamma\bigr]}{\delta\bar\Gamma(12,1'2')}=0
\label{dPhi_c/dbarGamma}
\end{align}
holds. Note also that $\bar\Phi_{\rm c}=\Phi_{\rm c}$ is satisfied,
as seen from Eqs.\ (\ref{Phi_L}) and (\ref{barPhi_c}).

In summary, the grand thermodynamic potential is now given as a functional of $(G,\Gamma)$ by
\begin{align}
\Omega[G,\Gamma]=&\,
\frac{\sigma}{\beta}{\rm Tr}\bigl[\ln(-\underline{G}_0^{-1}+\underline{\Sigma})+\underline{\Sigma}\,\underline{G}\bigr]+\Phi_{\rm MF}[G,\Gamma^{(0)}] 
\notag \\
&\,+\bar\Phi_{\rm c}\bigl[\bar\Gamma^{(0)},\bar\Gamma\bigr],
\label{barOmega}
\end{align}
which is stationary with respect to the variations of both $G$ and $\Gamma$, as given explicitly by Eqs.\ (\ref{Dyson}) and (\ref{dPhi_c/dbarGamma}).
The corresponding self-energy is defined by 
\begin{align}
\Sigma(1',1)\equiv -\sigma\beta\frac{\delta\bigl(\Phi_{\rm MF}+\bar\Phi_{\rm c}\bigr)}{\delta G(1,1')}
\label{Sigma2}
\end{align}
instead of Eq.\ (\ref{Sigma}).
Noting Eq.\ (\ref{dPhi_c/dbarGamma}), we can perform the differentiation of $\bar{\Phi}_{\rm c}$
with respect to $G$ by considering only the dependence through $\bar\Gamma^{(0)}$ given by Eq.\ (\ref{barGamma^(0)}).
We thereby obtain
\begin{align}
\Sigma(1',1)
=&\,\Sigma_{\rm MF}(1',1)
\notag \\
&\, \hspace{-10mm}+\frac{\sigma}{4}\Gamma^{(0)}(3'4',12)G(2,2')\Gamma(1'2',34)G(3,3')G(4,4')
\notag \\
&\, \hspace{-10mm}+\frac{\sigma}{4}\Gamma^{(0)}(1'2',34)G(3,3')G(4,4')\Gamma(3'4',12)G(2,2'),
\label{Sigma2a}
\end{align}
where $\Sigma_{\rm MF}(1',1)\!\equiv\!-\sigma \Gamma^{(0)}(1'2',12)G(2,2_+')$ is the mean-field self-energy and $\Gamma$ is defined in terms of $\bar\Gamma$ in the same way as Eq.\ (\ref{barGamma^(0)}).

This completes our basic formulation.
Specifically, Dyson's equation (\ref{Dyson}) with Eq.\ (\ref{Sigma2a}) and the stationarity condition $\delta \Omega/\delta\bar\Gamma=0$ 
constitute a set of self-consistent (i.e., nonlinear) equations for $G$ and $\Gamma$.
Substituting the solution into Eq.\ (\ref{barOmega}) yields the grand potential $\Omega$.

\subsection{Two-particle Green's function}

The formalism also enables us to obtain the two-particle Green's function,
\begin{subequations}
\label{G^II}
\begin{align}
G^{\rm II}(12,1'2')\equiv\langle \hat{T}_\tau \hat{\psi}(1)\hat{\psi}(2)\hat{\psi}^\dagger(2')\hat{\psi}^\dagger(1')\rangle,
\label{G^II-1}
\end{align}
where $\hat{T}_\tau$ is the time-ordering operator.\cite{LW60,Kita10}
Indeed, it is expressible as a sum of the disconnected and cumulant parts,\cite{AGD63}
\begin{align}
G^{\rm II}(12,1'2')=&\,
G(1,1')G(2,2')+\sigma G(1,2')G(2,1')
\notag \\
&\,\hspace{-13mm} -G(1,3')G(2,4')\Gamma(3'4',34)G(3,1')G(4,2'),
\label{G^II-2}
\end{align}
\end{subequations}
with $G(1,1')\equiv-\langle \hat{T}_\tau \hat{\psi}(1)\hat{\psi}^\dagger(1')\rangle$.
Equation (\ref{G^II-2}) can be regarded as defining the renormalized interaction vertex $\Gamma$ in such a way that
replacing $(G,\Gamma)$ by $(G_0,\Gamma^{(0)})$ 
yields the first-order perturbation result for $G^{\rm II}$.
It tells us that, once $G$ and $\Gamma$ are known, we can also calculate the two-particle Green's function.
Thus, the present formalism enables us to study two-particle or collective excitations simultaneously 
with the free energy and single-particle excitations
in a unified framework.

The validity of using the same $\Gamma$ in Eq.\ (\ref{G^II-2}) as that in Eq.\ (\ref{Sigma2a})
can be confirmed in terms of the interaction energy.
Specifically, it follows from Eqs.\ (\ref{H_int-1}) and (\ref{G^II-1}) that the interaction energy is expressible as
\begin{subequations}
\label{<H_int>}
\begin{align}
\langle \hat{H}_{\rm int}\rangle = &\,\frac{1}{4\beta} \Gamma^{(0)}(1'2',12)G^{\rm II}(12,1_+'2_+').
\label{<H_int>-1}
\end{align}
On the other hand, by
replacing $\hat{H}_{\rm int}$ by $\lambda \hat{H}_{\rm int}$, differentiating the resulting 
$\Omega(\lambda)$ in terms of $\lambda$, and setting $\lambda=1$,
we also obtain\cite{LW60,Kita10}
\begin{align}
\langle \hat{H}_{\rm int}\rangle =-\frac{\sigma}{2\beta}\Sigma(1',1)G(1,1_+').
\label{<H_int>-2}
\end{align}
\end{subequations}
By substituting Eqs.\ (\ref{G^II-2}) and (\ref{Sigma2a}) into Eqs.\ (\ref{<H_int>-1}) and (\ref{<H_int>-2}), respectively,
we obtain an identical expression for $\langle \hat{H}_{\rm int}\rangle$.
This agreement shows the consistency of the present formulation,
which generally cannot be reached by vertices obtained from
the Bethe--Salpeter equation using $\Gamma^{({\rm i})}\propto \delta^2\Phi/\delta G\delta G$
as the input called {\it irreducible vertex}.\cite{Baym62,Kita10,Kontani13}

\subsection{Comments\label{subsec:comments}}

Four comments are in order concerning the formalism.
First, the present definition of the renormalized vertex by Eq.\  (\ref{bGamma-def})
is advantageous over that of De Dominicis and Martin\cite{DDM64-2}
in terms of the following: (i) it is simple and algebraic without recourse to any topological structure of 
skeleton-type diagrams;
(ii) it reduces to $\Gamma^{(0)}$ instead of $-\Gamma^{(0)}$ in the weak-coupling limit.
The subsequent Legendre transformation from $\Gamma^{(0)}$ to $\Gamma$
can be performed straightforwardly from Eq.\ (\ref{Phi_L}) down to Eq.\ (\ref{barPhi_c}).

Second, the formalism is conserving, i.e., it satisfies the particle, momentum, and energy
conservation laws automatically. The proofs proceed in exactly the same way as those given for the Luttinger--Ward functional\cite{Baym62,Kita10} in terms of the self-energy in Eq.\ (\ref{Sigma2}).

Third, Eq.\ (\ref{barOmega}) can be extended straightforwardly to ordered phases such as superconductivity and ferromagnetism,
except for Bose--Einstein condensation;  the case of superconductivity will be presented below.
Note in this context that, for fermions, the stationarity condition of $\Omega$ with respect to the order parameter is naturally included
in the stationarity condition of Green's function in the static limit, which incorporates the possible emergence of spontaneous anisotropy or order.

Fourth, the formalism enables us to calculate thermodynamic, single-particle, and two-particle properties 
in a single consistent approximation scheme, 
in contrast to previous treatments where they have been
studied by adopting different approximations, 
e.g.,  the mean-field approximation for $(\Omega,G)$ and the random-phase approximation for $G^{\rm II}$.\cite{Anderson58,Moriya84}
Moreover, it is applicable to both normal and ordered phases on an equal footing 
because of the self-consistency procedure for calculating $(G,\Gamma)$, as already mentioned.
These two features may be regarded as definite advantages of the present formalism, especially 
for describing collective fluctuations near second-order transitions.
Note also that practical calculations of $(\Omega,G,G^{\rm II})$ can be performed in terms of $G$ and $\Gamma^{(0)}$ alone
without recourse to the Legendre transformation at all,
by using Eqs.\ (\ref{bGamma-def}) and (\ref{G^II}) additionally to study two particle properties such as transport coefficients.

\section{Examples for Fermions} 

We focus on the case of fermions with $\sigma=-1$ to present some examples 
of approximations, including one for superconductivity.

\subsection{Particle--particle scattering approximation}

First, we consider the $T$-matrix \cite{BK61,Baym62} or particle--particle scattering\cite{BS89} approximation 
suitably
generalized to incorporate the exchange process at each order.
This is one of the exceptional cases where Eq.\ (\ref{bGamma-def}) can be inverted analytically to express $\Gamma^{(0)}$ explicitly 
in terms of $\Gamma$ satisfying the antisymmetry requirement: $\Gamma(1'2',12)=-\Gamma(2'1',12)$.
Specifically, let us collect
the series of the third and fifth contributions in Fig.\ \ref{Fig1} up to infinite order besides that of the second-order diagram.
Accordingly, Eq.\ (\ref{Phi_c}) is approximated by
\begin{align}
\Phi_{\rm c}\bigl[\bar{\Gamma}^{(0)}\bigr]=&\,\frac{1}{\beta}{\rm Tr}\biggl[\ln \biggl(\underline{1}+\frac{1}{2} \underline{\bar{\Gamma}}^{(0)}\biggr)
- \frac{1}{2}\underline{\bar{\Gamma}}^{(0)}\biggr],
\label{Phi_c^pp}
\end{align}
where $\underline{1}$ and $\underline{\bar{\Gamma}}^{(0)}$ are matrices whose elements are given by
$(\underline{1})_{11',22'}\!\equiv\! \delta(1,2)\delta(1',2')$ and $(\underline{\bar{\Gamma}}^{(0)})_{11',22'}\!\equiv\! \bar{\Gamma}^{(0)}(11',22')$,
and the logarithm is defined by the power series of $\underline{\bar\Gamma}^{(0)}$.
The differentiation of Eq.\ (\ref{bGamma-def}) yields
\begin{align}
\underline{\bar\Gamma}= \underline{\bar\Gamma}^{(0)}\biggl( \underline{1}+\frac{1}{2}\underline{\bar\Gamma}^{(0)}\biggr)^{\!\!-1} .
\label{Gamma-Gamma^(0)-pp}
\end{align}
The relation can be inverted as
\begin{align}
\underline{\bar\Gamma}^{(0)} = \underline{\bar\Gamma}\,\biggl( \underline{1}-\frac{1}{2}\underline{\bar\Gamma}\biggr)^{\!\!-1}.
\label{Gamma-pp}
\end{align}
Substitution of Eq.\ (\ref{Gamma-pp}) into Eq.\ (\ref{Phi_L}) with Eq.\ (\ref{Phi_c^pp}) yields
\begin{align}
\Phi_{\rm L}\bigl[\bar\Gamma\bigr]\equiv \frac{1}{\beta}{\rm Tr}\biggl[-\ln \biggl(\underline{1}-\frac{1}{2} \underline{\bar{\Gamma}}\biggr)
-\frac{1}{2}\underline{\bar{\Gamma}}\biggr] .
\label{Phi_L^pp}
\end{align}
By comparing Eq.\ (\ref{Phi_L^pp}) with Eq.\ (\ref{Phi_c^pp}), one realizes that the vertex renormalization cannot be carried out by assigning the same sign and weight to each skeleton diagram as those in the bare perturbation expansion, contrary to the assumption in Ref.\ \onlinecite{DDM64-2}.
The transformation from Eq.\ (\ref{Phi_c}) to Eq.\ (\ref{barPhi_c}) may be regarded as an algebraic Legendre transformation
introduced without assuming the convexity of the original functional.\cite{JKM17}

The above results can be put into more familiar expressions
by replacing $\underline{\bar\Gamma}\rightarrow \underline{\Gamma}\,\underline{GG}$
with $(\underline{{\Gamma}})_{11',22'}\!\equiv\! {\Gamma}(11',22')$ and $\bigl(\underline{GG}\bigr)_{11',22'} \equiv G(1,2)G(1',2')$.

\subsection{Particle--hole scattering approximation}

Second, we consider the shielded potential\cite{BK61,Baym62} or particle--hole scattering approximation\cite{BS89}
suitably generalized to incorporate the exchange process at each order.
Specifically, let us collect
the fourth and sixth diagrams in Fig.\ \ref{Fig1} up to infinite order besides that of the second-order diagram.
Accordingly, Eq.\ (\ref{Phi_c}) is approximated by
\begin{align}
\Phi_{\rm c}\bigl[\bar{\Gamma}^{(0)}\bigr]=&\,\frac{1}{2\beta}{\rm Tr}\biggl[\ln \bigl(\underline{1}-\underline{\bar{\Gamma}}^{(0)}\bigr)
+\underline{\bar{\Gamma}}^{(0)}
+\frac{1}{4}\bigl(\underline{\bar{\Gamma}}^{(0)}\bigr)^2\biggr],
\label{Phi_c^ph}
\end{align}
where the matrix $\underline{\bar{\Gamma}}^{(0)}$ is now defined by $\langle 11'|\underline{\bar\Gamma}^{(0)}|22'\rangle\equiv \bar{\Gamma}^{(0)}(12',1'2)$.
Performing the differentiation of Eq.\ (\ref{bGamma-def}) yields
\begin{align}
\langle 11'|\underline{\bar\Gamma}|22'\rangle =&\, \langle 11'|\underline{\bar\Gamma}^{(0)}|22'\rangle
\notag \\
&\,
+\langle 11'|\bigl(\underline{\bar\Gamma}^{(0)}\bigr)^2\bigl( \underline{1}-\underline{\bar\Gamma}^{(0)}\bigr)^{\!\!-1}|22'\rangle 
\notag \\
&\,-\langle 12'|\bigl(\underline{\bar\Gamma}^{(0)}\bigr)^2\bigl( \underline{1}-\underline{\bar\Gamma}^{(0)}\bigr)^{\!\!-1}|21'\rangle.
\label{Gamma-Gamma^(0)-ph}
\end{align}
In general, the equation can only be inverted numerically to write $\Gamma^{(0)}$ in terms of $\Gamma$
owing to the presence of the exchange contribution given by the third term on the right-hand side.

On the other hand, we can express $\Omega$ analytically in terms of a renormalized vertex $\Gamma^{({\rm ph})}$ without 
the antisymmetry. Specifically, we introduce $\Gamma^{({\rm ph})}(11',22')\equiv 
\langle 11'|\underline{\Gamma}^{({\rm ph})}|22'\rangle$ by performing the differentiation of Eq.\ (\ref{bGamma-def}) 
without taking the antisymmetry of ${\bar\Gamma}^{(0)}$ into account to obtain
\begin{align}
\underline{\bar\Gamma}^{({\rm ph})} =\underline{\bar\Gamma}^{(0)}
+2\bigl(\underline{\bar\Gamma}^{(0)}\bigr)^2\bigl( \underline{1}-\underline{\bar\Gamma}^{(0)}\bigr)^{\!\!-1}.
\label{Gamma-Gamma^(0)-ph2}
\end{align}
The relation can be inverted algebraically as
\begin{align}
\underline{\bar\Gamma}^{(0)} = -\frac{1}{2}\bigl(\underline{1}+\underline{\Gamma}^{({\rm ph})}\bigr)
+\frac{1}{2} \Bigl[\underline{1}+6\underline{\Gamma}^{({\rm ph})}+\bigl(\underline{\Gamma}^{({\rm ph})}\bigr)^2\Bigr]^{\frac{1}{2}}.
\label{Gamma-ph}
\end{align}
The functional $\Phi_{\rm L}[\bar\Gamma^{({\rm ph})}]$ can be constructed
by substituting Eq.\ (\ref{Gamma-ph}) into Eq.\ (\ref{Phi_L}) with Eq.\ (\ref{Phi_c^ph}).
The results can be put into more familiar expressions
by the replacement $\underline{\bar\Gamma}^{({\rm ph})}\rightarrow \underline{\Gamma}^{({\rm ph})}\underline{GG}^{\rm T}$
with $\bigl(\underline{GG}^{\rm T}\bigr)_{11',22'} \equiv G(1,2)G(2',1')$.
The resulting functional $\bar\Phi[G,\Gamma^{({\rm ph})}]$ corresponds to the Luttinger--Ward $\Phi$ functional given in terms of $U$ instead of $\Gamma^{(0)}$.
It remains to be clarified how neglecting the antisymmetry requirement affects various results.
In this context, it is definitely necessary for calculating Eq.\ (\ref{G^II-2}) to substitute 
 $\frac{1}{2}[\Gamma^{({\rm ph})}(3'4',34)-\Gamma^{({\rm ph})}(4'3',34)]$ into $\Gamma(3'4',34)$
 so as to reproduce the antisymmetry of $G^{\rm II}$ appropriately.

\subsection{Fluctuation exchange approximation}

Third, we consider the fluctuation exchange (FLEX) approximation\cite{BS89} of collecting
the third to sixth diagrams in Fig.\ \ref{Fig1} up to infinite order besides that of the second-order diagram.
There are two ways of performing the renormalization. 

The first one is to solve Eq.\ (\ref{bGamma-bGamma^(0)}) in the FLEX approximation to obtain a single renormalized $\bar\Gamma$.
This approach is advantageous in that the mixing between the particle--particle and particle--hole processes is
naturally incorporated, but the approach will also be numerically much more demanding.

The second one neglects the antisymmetry requirement and
introduces two kinds of renormalized vertices corresponding the particle--particle and particle--hole channels.
Specifically, we introduce matrices $\underline{1}$, $\underline{\bar{\Gamma}}^{(0)}$,  and $\underline{C}$ through
\begin{subequations}
\begin{align}
(\underline{1})_{11',22'}\equiv &\,
\begin{bmatrix}\delta(1,2)\delta(1',2') & 0 \\
0 & \delta(1,2)\delta(1',2') 
\end{bmatrix},
\\
(\underline{\bar{\Gamma}}^{(0)})_{11',22'}\equiv &\,
\begin{bmatrix}\frac{1}{2}\bar{\Gamma}^{(0)}(11',22') & 0 \\
0 & -\bar{\Gamma}^{(0)}(12',1'2)
\end{bmatrix},
\\
(\underline{C})_{11',22'}\equiv &\,
\begin{bmatrix}2\delta(1,2)\delta(1',2') & 0 \\
0 & \delta(1,2)\delta(1',2') 
\end{bmatrix},
\end{align}
\end{subequations}
to incorporate the two 
channels independently.
Equation (\ref{Phi_c}) is then approximated by
\begin{align}
\Phi_{\rm c}\bigl[{\bar{\Gamma}}^{(0)}\bigr]=&\,\frac{1}{2\beta}{\rm Tr}\,\underline{C}
\biggl[\ln \bigl(\underline{1}+\underline{\bar{\Gamma}}^{(0)}\bigr)
-\underline{\bar{\Gamma}}^{(0)}
+\frac{1}{3}\bigl(\underline{\bar{\Gamma}}^{(0)}\bigr)^2\biggr] .
\label{Phi_c^FLEX}
\end{align}
The factor $\frac{1}{3}$ originates from our specific choice of dividing 
the second-order process into the particle--particle and particle--hole channels with the ratio 1:2, 
but note that there is arbitrariness in how the ratio is chosen.
Since Eq.\ (\ref{Phi_c^FLEX}) is diagonal in the particle--particle and particle--hole indices, we can perform the differentiation of 
Eq.\ (\ref{bGamma-bGamma^(0)}) easily to obtain
\begin{align}
\underline{\bar\Gamma}=&\,\begin{bmatrix}\underline{\bar\Gamma}^{({\rm pp})} & \underline{0} \\ 
\underline{0} & \underline{\bar\Gamma}^{({\rm ph})}
\end{bmatrix},
\label{Gamma-Gamma^(0)-FLEX}
\end{align}
where $\underline{\bar\Gamma}^{({\rm pp})}$ and $\underline{\bar\Gamma}^{({\rm ph})}$ are defined in terms of Eqs.\ (\ref{Gamma-Gamma^(0)-pp}) and  (\ref{Gamma-Gamma^(0)-ph2}), respectively, by subtracting $\frac{2}{3}\underline{\bar\Gamma}^{(0)}$ and $\frac{1}{3}\underline{\bar\Gamma}^{(0)}$; note the difference in the definitions of $\underline{\bar\Gamma}^{(0)}$ between them. Equation (\ref{Gamma-Gamma^(0)-FLEX}) can be used to express $\underline{\bar\Gamma}^{(0)}$ as a functional of $\underline{\bar\Gamma}$. Substituting it into Eq.\ (\ref{Phi_L}) with Eq.\ (\ref{Phi_c^FLEX}), we obtain $\Phi_{\rm L}[\bar\Gamma]$.

\subsection{FLEX-S approximation for superconductivity}

The FLEX approximation\cite{BS89} can be generalized concisely into the FLEX-S approximation
for describing superconductivity \cite{Kita11}. 
We will discuss the vertex renormalization in the FLEX-S approximation without requiring antisymmetry.

Let us express Eq.\ (\ref{H_int-1})
in a symmetric form with respect to $(\hat{\psi},\hat{\psi}^\dagger)\!\equiv\! (\hat{\psi}_{1},\hat{\psi}_2)$ as\cite{Kita11}
\begin{align}
\hat{H}_{\rm int}=&\,\frac{1}{4! \beta}\Gamma^{(0)}_{i'i,j'j}(1'1,2'2)
\notag \\
&\,\times \hat{\cal N} \hat{\psi}_{3-i'}(1')\hat{\psi}_{i}(1)
\hat{\psi}_{3-j'}(2') \hat{\psi}_{j}(2),
\label{H_int-2}
\end{align}
where $\hat{\cal N}$ is the normal-ordering operator of placing creation operators to the left,
and $\Gamma^{(0)}_{i'i,j'j}(1'1,2'2)$ with  $i,j=1,2$ is defined in terms of $\Gamma^{(0)}(1'2',12)$ in Eq.\ (\ref{Gamma^(0)}) by
\begin{figure}[t]
\begin{center}
\includegraphics[width=0.4\textwidth]{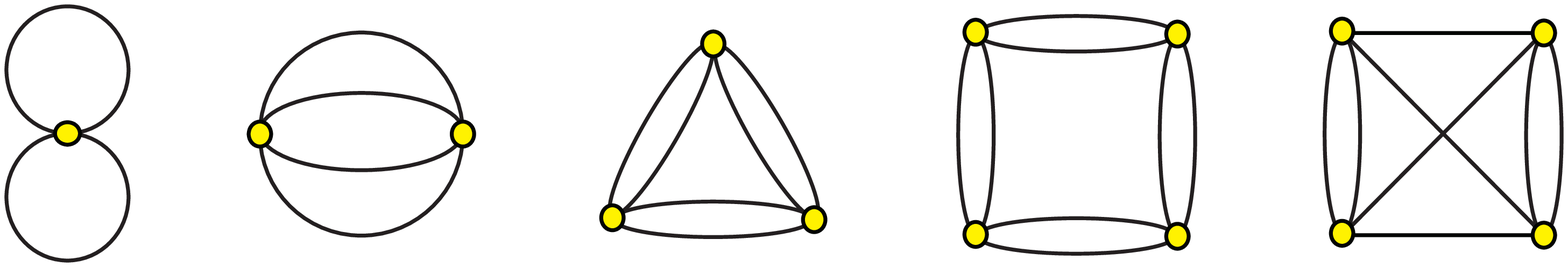}
\caption{(Color online) Diagrammatic expression of $\Phi$ up to the fourth order in terms of the vertex of Eq.\ (\ref{Gamma-mat}),
which is denoted by a small circle.
\label{Fig2}}
\end{center}
\end{figure}
\begin{align}
\Gamma^{(0)}_{i'i,j'j}(1'1,2'2)
\equiv &\, \delta_{ij}\delta_{i'i}\delta_{j'j}\Gamma^{(0)}(1'2',12)
\notag \\
&\,-\delta_{i,3-j}\delta_{i'i}\delta_{j'j}\Gamma^{(0)}(1'2,12')
\notag \\
&\, -\delta_{i,3-j}\delta_{i'j}\delta_{j'i}\Gamma^{(0)}(1'1,2'2),
\label{Gamma-mat}
\end{align}
having the symmetry
$\Gamma^{(0)}_{i'i,j'j}(1'1,2'2)\!=\!\Gamma^{(0)}_{j'j,i'i}(2'2,1'1)\!=\!-\Gamma^{(0)}_{i'j,j'i}(1'2,2'1)
\!=\!-\Gamma^{(0)}_{i'i,3-j,3-j'}(1'1,22')$.
The equivalence of Eqs.\ (\ref{H_int-1}) and (\ref{H_int-2}) can be seen easily by substituting Eq.\ (\ref{Gamma-mat})
into the latter and rearranging the resulting expression in the normal order to remove $\hat{\cal N}$.
The advantage of Eq.\ (\ref{H_int-2}) lies in the equivalence of the four field operators.
Indeed, by using the symmetry of $\Gamma^{(0)}_{i'i,j'j}$ and the anti-commutation relations of the field operators under $\hat{\cal N}$, we can place 
each of $\hat\psi_{3-j'}(2')$ and $\hat\psi_j(2)$ right next to $\hat\psi_{3-i'}(1')$ and transform the resulting expression 
into the same form as Eq.\ (\ref{H_int-2}) through a change of variables.
This equivalence enables us to perform the perturbation expansion of $\Phi$ in $\hat{H}_{\rm int}$ concisely
in terms of the Feynman diagrams of Fig.\ \ref{Fig2} without arrows,\cite{Kita11}
where an additional summation at each vertex over every internal index $i=1,2$ is implied.
To be more specific, we can associate $(\hat\psi_i,\hat\psi_{3-i})$ in Eq.\ (\ref{H_int-2}) with $(\hat\psi,\hat\psi^\dagger)$ of the normal state
to write down a single contribution analytically for each Feynman diagram 
in terms of Green's function,
\begin{equation} 
G_{ii'}(1,1')\equiv - \bigl< \hat{T}_\tau \hat{\psi}_{i}(1)\hat{\psi}_{3-i'}(1')\bigr> .
\label{G_ij}
\end{equation}
and subsequently multiply it by a combinatorial factor originating from the equivalence of the four field operators. 
By introducing matrices $\underline{1}$, $\underline{\Gamma}^{(0)}$, and $\underline{GG}^{\rm T}$ as
\begin{subequations}
\begin{align}
\langle 1'1_{i'i}|\underline{1}|2'2_{j'j}\rangle\equiv&\, \delta_{ij}\delta_{i'j'}\delta(1,2)\delta(1',2'),
\\
\langle 1'1_{i'i}|\underline{\Gamma}^{(0)}|2'2_{j'j}\rangle\equiv&\, \Gamma^{(0)}_{i'i,jj'}(1'1,22'),
\\
\langle 1'1_{i'i}|\underline{GG}^{\rm T}|2'2_{j'j}\rangle\equiv&\, G_{i'j'}(1',2')G_{ji}(2,1),
\end{align}
\end{subequations}
we can thereby express the $\Phi$ functional in the FLEX-S approximation as
$\Phi=\Phi_{\rm MF}+\Phi_{\rm c}$ with\cite{Kita11}
\begin{subequations}
\label{Phi-S}
\begin{align}
\Phi_{\rm MF}=&\,-\frac{1}{8\beta}{\rm Tr}\,\underline{\Gamma}^{(0)}\underline{GG}^{\rm T},
\\
\Phi_{\rm c}=&\, \frac{1}{2\beta}{\rm Tr}\biggl[\ln\biggl(\underline{1}-\frac{1}{2}\underline{\Gamma}^{(0)}\underline{GG}^{\rm T}\biggr)
+\frac{1}{2}\underline{\Gamma}^{(0)}\underline{GG}^{\rm T}
\notag \\
&\, +\frac{1}{12}\bigl(\underline{\Gamma}^{(0)}\underline{GG}^{\rm T}\bigr)^2\biggr].
\label{Phi_c-S}
\end{align}
\end{subequations}
See Eq.\ (28) of Ref.\ \onlinecite{Kita11} and note the differences of factor 2 in the definition of 
Eq.\ (\ref{Gamma-mat}) and the arrangement of arguments in Eq.\ (\ref{Gamma^(0)}) from Eqs.\ (20b) and (16) in Ref.\ \onlinecite{Kita11},
respectively. 
Note that Eq.\ (\ref{Phi_c-S}) appropriately reduces to Eq.\ (\ref{Phi_c^FLEX}) in the normal-state limit of
$G_{12}=G_{21}=0$.

By decomposing $\underline{G}=\underline{G}^{\frac{1}{2}}\underline{G}^{\frac{1}{2}}$
with $\langle 1_i|\underline{G}| 1'_{i'}\rangle \equiv G_{ii'}(1,1')$, we can express Eq.\ (\ref{Phi_c-S})
as a functional of 
\begin{align}
\underline{\bar\Gamma}^{(0)}\equiv \underline{G}^{\frac{1}{2}}\bigl(\underline{G}^{\rm T}\bigr)^{\frac{1}{2}}\underline{\Gamma}^{(0)}
\underline{G}^{\frac{1}{2}}\bigl(\underline{G}^{\rm T}\bigr)^{\frac{1}{2}}
\end{align}
alone by the replacement $\underline{\Gamma}^{(0)}\underline{GG}^{\rm T}\rightarrow \underline{\bar\Gamma}^{(0)}$.
Equation (\ref{bGamma-def}) is now changed into $\bar\Gamma=-4! \beta \delta\Phi_{\rm c}/\delta\bar\Gamma^{(0)}$, 
and the substitution of
Eq.\ (\ref{Phi_c-S}) into its right-hand side yields
\begin{align}
\underline{\bar\Gamma}= \underline{\bar\Gamma}^{(0)}+\frac{3}{2}\bigl(\underline{\bar\Gamma}^{(0)}\bigr)^2\biggl( \underline{1}-\frac{1}{2}\underline{\bar\Gamma}^{(0)}\biggr)^{\!\!-1} .
\label{Gamma-Gamma^(0)-FLEXS}
\end{align}
The relation can be inverted algebraically as
\begin{align}
\underline{\bar\Gamma}^{(0)} = -\frac{1}{2}\biggl(\underline{1}+\frac{1}{2}\underline{\Gamma}\biggr)
+\frac{1}{2} \biggl[\underline{1}+5\underline{\Gamma}+\frac{1}{4}\bigl(\underline{\Gamma}\bigr)^2\biggr]^{\frac{1}{2}}.
\label{Gamma-FLEXS}
\end{align}
We then perform the Legendre transformation
\begin{align}
\Phi_{\rm L}[\bar\Gamma]\equiv \frac{1}{4!\beta}{\rm Tr}\,\underline{\bar\Gamma}^{(0)}
\underline{\bar\Gamma}+\Phi_{\rm c}[\bar\Gamma^{(0)}],
\end{align}
which satisfies $\bar\Gamma^{(0)}=4! \beta \delta\Phi_{\rm L}/\delta\bar\Gamma$,
and introduce
\begin{align}
\bar\Phi_{\rm c}[\underline{\bar\Gamma}^{(0)},\underline{\bar\Gamma}]\equiv 
-\frac{1}{4!\beta}{\rm Tr}\,\underline{\bar\Gamma}^{(0)}
\underline{\bar\Gamma}+\Phi_{\rm L}[\bar\Gamma] .
\label{Phi-S2}
\end{align}
The grand thermodynamic potential $\Omega=\Omega[\underline{G},\underline{\Gamma}]$ 
is given in terms of Eq.\ (\ref{Phi-S2}) by
\begin{align}
\Omega=-\frac{1}{2\beta}{\rm Tr}\,\bigl[\ln\bigl(-\underline{G}_0^{-1}+\underline{\Sigma}\bigr)+\underline{G}\,\underline{\Sigma}\bigr]
+\Phi_{\rm MF}+\bar\Phi_{\rm c}.
\label{Omega-super}
\end{align}
See Eq.\ (11) of Ref.\ \onlinecite{Kita11}.
It satisfies $\delta\Omega/\delta \underline{G}\!=\!\underline{0}$ and $\delta\Omega/\delta \underline{\Gamma}\!=\!\underline{0}$.

\section{Concluding Remarks}

We have developed a formalism of calculating the grand  thermodynamic potential $\Omega$, one-particle Green's function $G$, 
and two-particle Green's function $G^{\rm II}$ in a unified consistent framework by expressing $\Omega$
as a functional of $G$ and the renormalized interaction vertex $\Gamma$.
The key result  is Eq.\ (\ref{barOmega}) for $\Omega$, 
which satisfies the stationarity conditions $\delta\Omega/\delta G=0$ and $\delta\Omega/\delta\Gamma=0$.
Using the solution of the coupled equations, we obtain $G$ by Eq.\ (\ref{Dyson}) in terms of the self-energy given by Eq.\ (\ref{Sigma2a}), 
$G^{\rm II}$ by Eq.\ (\ref{G^II-2}), and $\Omega$ by Eq.\ (\ref{barOmega}); this $\Omega$
naturally contains contributions of both single-particle and two-particle (i.e., independent and collective) excitations.
The differences from and the advantages of the present formalism over previous ones are summarized in Sect.\ \ref{subsec:comments}.
Functional (\ref{barOmega}) will also be useful in phenomenological studies of adopting some model form for $\Gamma$
to clarify collective-mode contributions to thermodynamic observables.
%, especially in extending the Fermi-liquid theory given in terms of single-particle excitations alone 
%to incorporate collective fluctuations on a well-defined microscopic basis.
Generalizing the formalism to describe Bose--Einstein condensates
with a finite average $\langle\hat\psi\rangle$ has yet to be performed.

A way to incorporate single-particle and collective excitations simultaneously into calculations of thermodynamic observables 
has been sought over many decades with no definite answer reached yet apparently. 
Back in the 1960, Doniach and Engelsberg\cite{DE67} expressed the interaction energy of nearly ferromagnetic Fermi liquids 
in terms of the magnetic susceptibility in the random-phase approximation obtained earlier by Izuyama {\it et al.},\cite{IKK63}
integrated it in terms of the coupling constant to extract the extra free energy due to spin fluctuations, 
and obtained a qualitative fit to the low-temperature specific heat of liquid $^3$He that exhibits a logarithmic non-Fermi-liquid behavior
(see Ref.\ \onlinecite{BP91} for more references on the subject).
The approach has also been adopted by Moriya\cite{Moriya84} in constructing the theory of itinerant electron magnetism,
where susceptibility is expressed in terms of several phenomenological parameters whose values can be extracted from experiments.
On the other hand, the logarithmic corrections to Landau's Fermi-liquid theory observed in various physical quantities have been
described alternatively by introducing the concept of {\it statistical quasiparticle energy} in the single-particle channel.\cite{BP91}
The extensive references on the subject given in Ref.\ \onlinecite{BP91} indicates the absence of any established microscopic framework 
for treating single-particle and collective excitations simultaneously and consistently for calculating thermodynamic observables such as 
specific heat. 
The situation was unchanged until today, as may be seen in a recent attempt to incorporate collective Cooper-pair fluctuations
into the Fermi liquid theory near superfluid transitions.\cite{LS22}
The present formalism, which can handle $(\Omega,G,G^{\rm II})$ microscopically and simultaneously in a unified consistent manner,
is expected to provide a firm basis for studying the two kinds of contributions to thermodynamic observables quantitatively.

\section*{Acknowledgment}
This work was supported by JSPS KAKENHI Grant Number JP20K03848.


\begin{thebibliography}{99}
\bibitem{BK61}G. Baym and L. P. Kadanoff, Phys. Rev. {\bf 124}, 287 (1961); see also, L. P. Kadanoff and G. Baym, {\it Quantum Statistical Mechanics} (W.A. Benjamin, New York, 1962).
\bibitem{Baym62}G. Baym, Phys. Rev. {\bf 127}, 1391 (1962).
\bibitem{BS89}N. E. Bickers and D. J. Scalapino, Ann. Phys. {\bf 193}, 206 (1989).
\bibitem{DDM64-1}C. De Dominicis and P. C. Martin, J. Math. Phys. {\bf 5}, 14 (1964).
\bibitem{DDM64-2}C. De Dominicis and P. C. Martin, J. Math. Phys. {\bf 5}, 31 (1964).
\bibitem{Stratotovich57}R. L. Stratonovich, Sov. Phys. Doklady {\bf 2}, 416 (1957).
\bibitem{Hubbard59}J. Hubbard, Phys. Rev. Lett. {\bf 3}, 77 (1959).
\bibitem{LW60}J. M. Luttinger and J. C. Ward, Phys. Rev. {\bf 118}, 1417 (1960).
\bibitem{Kita10}T. Kita, Prog. Theor. Phys. {\bf 123}, 581 (2010).
\bibitem{Kita15}T. Kita, {\it Statistical Mechanics of Superconductivity} (Springer, Tokyo, 2015), Chap.\ 3.
\bibitem{AGD63}A. A. Abrikosov, L. P. Gorkov, and I. E. Dzyaloshinski, {\it Methods of Quantum Field Theory in Statistical Physics}
(Dover, New York, 1963), p.\ 92.
\bibitem{Kontani13}H. Kontani, {\it Transport Phenomena in Strongly Correlated Fermi Liquids} (Springer-Verlag, Berlin, 2013).
\bibitem{Anderson58}P. W. Anderson, Phys. Rev. {\bf 112}, 1900 (1958).
\bibitem{Moriya84}T. Moriya, {\it Spin Fluctuations in Itinerant Electron Magnetism} (Springer-Verlag, Berlin, 1985).
\bibitem{JKM17}D. M. Jackson, A. Kempf, and A. H. Morales, J. Phys. A: Math. Theor. {\bf 50}, 225201 (2017).
\bibitem{Kita11}T. Kita, J. Phys. Soc. Jpn. {\bf 80}, 124704 (2011).
\bibitem{DE67}S. Doniach and S. Engelsberg, Phys. Rev. Lett. {\bf 17}, 750 (1966).
\bibitem{IKK63}T. Izuyama, D. J. Kim, and R. Kubo, J. Phys. Soc. Jpn. {\bf 18}, 1025 (1963).
\bibitem{BP91}G. Baym and C. Pethick, {\it Landau Fermi-Liquid Theory: Concepts and Applications} (John Wiley \& Sons, New York, 1991).
\bibitem{LS22}W.-T. Lin and J. A. Sauls, Prog. Theor. Exp. Phys. {\bf 2022}, 033102.
%\bibitem{Bickers04}N. E. Bickers, in {\it Theoretical Methods for Strongly Correlated Electrons} (Springer-Verlag, New York, 2004), ed. by D. S\'en\'echal, A.-M. Tremblay, C. Bourbonnais, p. 237.


\end{thebibliography}
\end{document}